\documentclass[11pt,letter,reqno]{amsart}
\usepackage[latin9]{inputenc}
\usepackage{amstext}
\usepackage{amsthm}
\usepackage{amssymb}
\usepackage{graphicx}
\usepackage{comment}
\usepackage{amsmath}
\usepackage{breqn}
\makeatletter
\numberwithin{equation}{section}
\numberwithin{figure}{section}
\theoremstyle{plain}

 \theoremstyle{remark}

\newtheorem{theorem}{Theorem}[section]

\newtheorem{remark}[theorem]{Remark}

\usepackage{amsthm}
\usepackage{color}
\usepackage{color}
\usepackage[dvipsnames]{xcolor}
\usepackage{subcaption}
\usepackage{enumerate}
\usepackage[draft]{todonotes}
\usepackage[american]{babel}
\usepackage{geometry}

\newcommand{\ri}{\mathrm{i}}
\newcommand{\eig}{\mathrm{eig}}

\newcommand{\sign}{\mathrm{sign}}
\newcommand{\erf}{\mathrm{erf}}

\makeatother

  \providecommand{\remarkname}{Remark}
\providecommand{\theoremname}{Theorem}

\begin{document}

\title[Hydrodynamic Manifold Kinetic Equation]{Non-Local Hydrodynamics as a Slow Manifold for the One-Dimensional Kinetic Equation}

\author{Florian Kogelbauer}
\address{Institute for Mechanical Systems, ETH Z\"{u}rich, Leonhardstrasse 21, 8092 Z\"{u}rich, Switzerland}
\email{floriank@ethz.ch}

\maketitle
\begin{abstract}
	We prove an explicit, non-local hydrodynamic closure for the linear one-dimensional kinetic equation independent on the size of the relaxation time. We compare this dynamical equation to the local approximations obtained from the Chapman--Enskog expansion for small relaxation times. Our results rely on the spectral theory of Jacobi operators with rank-one perturbations.
\end{abstract}

\section{Introduction}

One of the main problems of statistical mechanics consists of the hydrodynamic closure of Boltzmann(-type) equations. Given the evolution equation for the distribution function, one seeks to recover the classical equations of fluid dynamics, such as the Euler equation or the Navier--Stokes equation, by formally expanding in a small parameter $\varepsilon$, the Knudsen number. This widely used technique, i.e., expanding the unknown distribution function in a formal Taylor series in $\varepsilon$, is called Chapman--Enskog (CE) series. In this fashion, one obtains an approximate closure in terms of the hydrodynamic variables at each order or $\varepsilon$. For a general treatment of the hydrodynamic closure (in the context of Hilbert's $6^{th}$ problem) we refer to \cite{gorban2014hilbert}.\\
The smallness assumption of the Knudsen number, however, limits the range of physical applicability of the models, thus prompting the quest for hydrodynamic models beyond the first few terms of the CE expansion. Even more dramatically, the Burnett and super-Burnett approximations of the Boltzmann equation (higher-order approximations in $\varepsilon$)  exhibit non-physical behavior, such as unboundedness in wave number and even instability, see \cite{bobylev1982chapman}.\\
For some model equations, however, the CE series can be summed explicitly, see \cite{gorban1996short} and \cite{karlin2014non}, leading to a non-local (in the sense of integral operators) hydrodynamic closure. The main idea is to assume that the hydrodynamic variables form an invariant manifold to which the overall dynamics converges sufficiently fast. For a detailed and general treatment of invariant manifolds in the context of physical dynamics, we refer to \cite{gorban2005invariant}. From a dynamical system point-of-view, this is justified if the the time-scales are well separated (slow-fast dynamics), leading to an exponentially fast convergence to the slow manifold. For simple systems, the slow-fast point of view is consistent with the exact summation of the CE series \cite{Kogelbauer2019}.\\
In this paper, we are concerned with the one-dimensional kinetic equation
\begin{equation}\label{kinetic}
	f_t=-vf_x-\frac{1}{\tau}(f-f^{eq}),
\end{equation}
for the relaxation time $\tau$ and some equilibrium function $f^{eq}$ (for details, see the following section). Let $\rho=\int_{\mathbb{R}}f\,dv$ denote the locally conserved density. 
We prove that equation \eqref{kinetic} admits a non-local (exact) hydrodynamic closure of the form
\begin{equation}\label{closedrho}
\frac{\partial}{\partial t}\rho(x,t)=\frac{1}{2\pi}\int_{\mathbb{R}}\lambda^*(k,\tau)\hat{\rho}(k,t)e^{\ri k x}\, dx,
\end{equation}
where $\lambda^*(k,\tau)$ is related to the implicit solution of a transcendental equation and $\hat{\rho}$ is the Fourier transform of $\rho$. From equation \eqref{closedrho}, we can recover the classical CE expansion of the kinetic equation by Taylor expanding the multiplyer $\lambda^*$ in $\tau$.\\
This is consistent with the non-perturbative analytical techniques developed in \cite{karlin2014non}. The approach perused in this paper based on spectral theory of Jacobi operators with rank-one perturbations gives an explicit (up to the solution of a transcendental equation) description of the non-local hydrodynamic closure. Furthermore, we provide  a formula for the critical wave number beyond which the slow mode ceases to exist.

\section{Hydrodynamics  as Slow Manifolds}
\subsection{Setup}
Consider the one-dimensional kinetic equation for the distribution function $f:\mathbb{R}^2\times [0,\infty), \, (x,v,t)\mapsto f(x,v,t)$,
\begin{equation}\label{maineq}
\frac{\partial f}{\partial t}=-v\frac{\partial f}{\partial x} -\frac{1}{\tau}(f-f^{eq}),
\end{equation}
where 
\begin{equation}
f^{eq}(x,v,t)=\rho(x,t)\sqrt{\frac{m}{2\pi k_BT}}e^{-\frac{mv^2}{2k_BT}},
\end{equation}
is the Maxwellian equilibrium function and
\begin{equation}
\rho(x,t)=\int_{\mathbb{R}}f(x,v,t)\, dv,
\end{equation}
is the locally conserved density. 
Without loss of generality, we can set
\begin{equation}
\frac{m}{k_BT}=1,
\end{equation}
by re-scaling the density function $f$ in $x$ and $v$. To analyze the slow dynamics of equation \eqref{maineq} in more detail, we will expand the density $f$ in in velocity space, using orthogonal polynomials, as well as in position space, using the Fourier transform.\\ Hermite polynomials constitute a natural basis for the expansion of $f$ in terms of momenta, i.e., in velocity space, see e.g. \cite{grad1949kinetic} and \cite{colosqui2010high}. We will use the probabilist's' version, defined as
\begin{equation}
H_n(v)=(-1)^ne^{\frac{v^2}{2}  }\left(\frac{d}{dv}\right)^ne^{-\frac{v^2}{2}}.
\end{equation}
With this convention, the first four Hermite polynomials are given as
\begin{equation}
\begin{split}
H_0(v)&=1,\\
H_1(v)&=v,\\
H_2(v)&=v^2-1,\\
H_3(v)&=v^3-3v.
\end{split}
\end{equation}
The set $\{H_n\}_{n\in\mathbb{N}}$ forms a complete orthogonal basis of the Hilbert space $\mathcal{H}=L^2\left(\mathbb{R}, e^{-\frac{v^2}{2}}dv \right)$ with the inner product
\begin{equation}
\langle f, g\rangle_{\mathcal{H}}=\int_{\mathbb{R}} f(v) g^*(v) e^{\frac{-v^2}{2}}\,  dv.
\end{equation}
Indeed, integration by parts shows that
\begin{equation}
\langle H_n, H_m\rangle_{\mathcal{H}}= \sqrt{2\pi}n! \delta_{n,m},
\end{equation}
for any $n,m\in\mathbb{N}$, where
\begin{equation}\label{delta}
\delta_{n,m}=\begin{cases} 0 &\text{ if } m\neq 0,\\
 1 &\text{ if } n=m,
\end{cases}
\end{equation}
is Dirac's delta. Let
\begin{equation}\label{normHermite}
\hat{H}_n(v):=\frac{1}{\sqrt{n!\sqrt{2\pi}}}H_n(v),\quad n\geq 0,
\end{equation}
be the normalized sequence of Hermite polynomials.  
We expand the density $f$ as
\begin{equation}
f(x,v,t)=e^{-\frac{v^2}{2}}\sum_{n=0}^{\infty} f_n(x,t) \hat{H}_n(v),
\end{equation}
with
\begin{equation}
f_n(x,t)=\int_{\mathbb{R}} f(x,v,t) \hat{H}_n(v)\,  dv,
\end{equation}
and note that
\begin{equation}
f_{eq}(x,v,t)=\frac{1}{\sqrt{2\pi}}\int_{\mathbb{R}}f(x,v,t)\, dv e^{-\frac{v^2}{2}}=\frac{1}{(2\pi)^{\frac{1}{4}}}f_0(x,t)e^{-\frac{v^2}{2}}=f_0(x,t)e^{-\frac{v^2}{2}}\hat{H}_0.
\end{equation}
The Hermite polynomials obey the Favard--Shohat recurrence formula
\begin{equation}\label{recurrence}
vH_n(v)=H_{n+1}(v)+nH_{n-1}(v), \quad n\geq 0,
\end{equation}
where $H_{-1}=0$, while the normalized Hermite polynomials obey the recurrence formula
\begin{equation}\label{recurrencenorm}
v\hat{H}_n(v)=\sqrt{n+1}\hat{H}_{n+1}(v)+\sqrt{n}\hat{H}_{n-1}(v), \quad n\geq 0,
\end{equation}
where $\hat{H}_{-1}=0$, which follows from dividing equation \eqref{recurrence} by $\sqrt{n!\sqrt{2\pi}}$.\\
Taking an inner product of equation \eqref{maineq} with $\hat{H}_n$, we calculate for $n\geq 0$:
\begin{equation}\label{innerprod1}
\begin{split}
\langle f_t,\hat{H}_n\rangle_{\mathcal{H}}&=\langle -vf_x -\frac{1}{\tau} (f-f^{eq}), \hat{H}_n\rangle\\
&=-\sum_{m=0}^\infty\left( \frac{\partial f_m}{\partial x}\langle v\hat{H}_m,\hat{H}_n\rangle\right)-\frac{1}{\tau}(f_n-f_0\delta_{0,n})\\
&=-\sum_{m=0}^\infty\left( \frac{\partial f_m}{\partial x}\langle \sqrt{m+1}\hat{H}_{m+1}+\sqrt{m}\hat{H}_{m-1},H_n\rangle\right)-\frac{1}{\tau}(f_n-f_0\delta_{0,n})\\\
&=-\left(\sqrt{n} \frac{\partial f_{n-1}}{\partial x}+\sqrt{n+1}\frac{\partial f_{n+1}}{\partial x}+\frac{1}{\tau}(f_n-f_0\delta_{0,n})\right),\\
\end{split}
\end{equation}
for $n\geq 1$. It follows from \eqref{innerprod1} that equation \eqref{maineq} is equivalent to the following sequence of equations for the momenta of $f$:
\begin{equation}\label{eqmoment}
\frac{\partial f_n}{\partial t}=-\sqrt{n}\frac{\partial f_{n-1}}{\partial x}-\frac{1}{\tau}f_n-\sqrt{n+1}\frac{\partial f_{n+1}}{\partial x}+\frac{1}{\tau}f_0\delta_{0,n},\quad n\geq 0.
\end{equation}
Taking the Fourier transform in the spatial variable of equation \eqref{eqmoment} and denoting
\begin{equation}
\hat{f}_n(k,t)=\int_{\mathbb{R}}f_n(x,t) e^{-\ri xk}\, dx,\quad n\geq 0,
\end{equation}
we arrive at the infinite-dimensional system
\begin{equation}\label{eqmomentFourier}
\frac{\partial \hat{f}_n}{\partial t}=-\ri k\sqrt{n} \hat{f}_{n-1}-\frac{1}{\tau}\hat{f}_n-\ri k\sqrt{n+1}\hat{f}_{n+1}+\frac{1}{\tau}\hat{f}_0\delta_{0,n},\quad n\geq 0,
\end{equation}
We bundle the momenta in a sequence $\mathbf{F}=(\hat{f}_0,\hat{f}_1,\hat{f}_2,...)$ and interpret equation \eqref{eqmomentFourier}  as an operator equation
\begin{equation}\label{eqF}
\mathbf{F}_t=\mathbf{T}(k)\mathbf{F},
\end{equation}
for the symmetric infinite-dimensional matrix
\begin{equation}\label{T}
\mathbf{T}(k)=\left(\begin{matrix}
0  & -\ri k & 0 & 0 & 0 & 0 &\ldots \\
-\ri k & -\frac{1}{\tau} & - \ri k\sqrt{2} & 0 & 0 & 0 &\dots \\
0 &  - \ri k\sqrt{2}  & -\frac{1}{\tau} & - \ri k\sqrt{3}  & 0 & 0 & \dots\\
0 & 0 &  - \ri k\sqrt{3} & -\frac{1}{\tau} & - \ri k\sqrt{4} & 0 & \ldots\\
0 & 0 & 0 &  - \ri k\sqrt{4}  & -\frac{1}{\tau} &  - \ri k\sqrt{5} & \dots\\
\vdots & \vdots & \vdots & \vdots & \vdots & \vdots & \ddots
\end{matrix}\right),
\end{equation}
depending on the wave number $k$.\\

\subsection{Spectral Analysis of the Operator $\mathbf{T}(k)$}
Let us analyze the spectrum of the operator \eqref{T} in dependence on the wave number $k$ in more detail. The operator \eqref{T} is a Jacobi operator on the space of square-summable sequences, cf. \cite{teschl2000jacobi}. Let $\sigma\Big(\mathbf{T}(k)\Big)$ denote the spectrum of $\mathbf{T}(k)$.\\
The resolvent of $\mathbf{T}$ is given by the solution operator of the second-order linear recurrence
\begin{equation}\label{resolvent}
\begin{split}
-\ri k \sqrt{n} \hat{f}_{n-1}-\left(\frac{1}{\tau}+\lambda\right)\hat{f}_n-\ri k \sqrt{n+1}\hat{f}_{n+1}&=\eta_n,\quad n\geq 1,\\
-\ri k \hat{f}_1-\lambda\hat{f}_0&=\eta_0,
\end{split}
\end{equation}
for $\lambda\in\mathbb{C}\setminus\sigma\Big(\mathbf{T}(k)\Big)$.\\
First, let us have a look at the case $k= 0$. System \eqref{resolvent} simplifies to
\begin{equation}\label{resolvent0}
\begin{split}
-\left(\frac{1}{\tau}+\lambda\right)\hat{f}_n&=\eta_n,\quad n\geq 1,\\
-\lambda\hat{f}_0&=\eta_0,
\end{split}
\end{equation}
which, in turn, implies that 
\begin{equation}
\sigma\Big(\mathbf{T}(0)\Big)=\left\{-\frac{1}{\tau},0\right\}.
\end{equation}
Here, the spectrum is a pure point spectrum and the corresponding eigenspaces are given as
\begin{equation}
\begin{split}
\eig\left(0\right)=\text{span}\Big((1,0,0,0...)\Big),\quad
\eig\left(-\frac{1}{\tau}\right)=\text{span}\Big((1,0,0,0...)\Big)^{\perp}.
\end{split}
\end{equation}

\begin{figure}
	\includegraphics[scale=0.6]{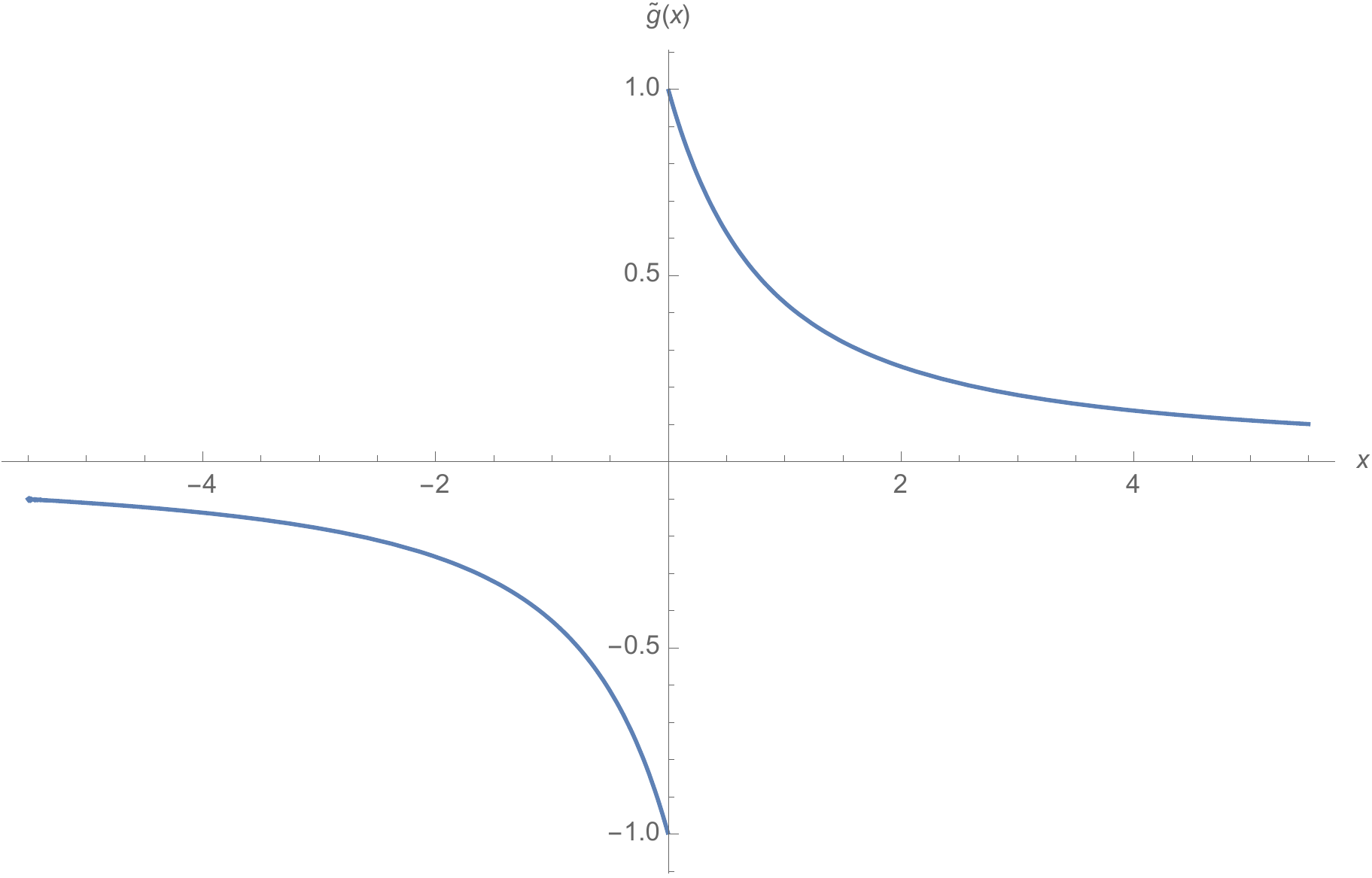}\caption{The functions $\tilde{g}(x)=e^{x^2}\left(\sign (x)-\erf(x)\right)$ derived from the diagonal Green's function evaluated at imaginary numbers.}
	\label{Green}
\end{figure}

Now, let us have a look at the case $k\neq 0$. Since the parameter $k$ enters the operator $\mathbf{T}$ through the highest finite difference, the perturbation for non-zero $k$ is singular. The eigenstructure of $\mathbf{T}(0)$ is highly degenerate, as $\mathbf{T}(0)$ admits an eigenvector of infinite multiplicity. According to analytic perturbation theory, see e.g.\cite{hislop2012introduction}, we do not expect analyticity of the spectrum of $\mathbf{T}(k)$ as $k\to 0$. Later explicit calculation will, however, confirm that the eigenvalue above the essential spectrum can be Taylor expanded around $k=0$ or $\tau=0$ up to a pole of order one.\\
We can write the operator \eqref{T} as a sum
\begin{equation}\label{Tsplit}
\mathbf{T}(k)=-\ri k \mathbf{S}(k)-\frac{1}{\tau}\mathbf{I}+\frac{1}{\tau}\mathbf{R}_1,
\end{equation}
where $\mathbf{I}$ is the identity, the symmetric operator $\mathbf{S}$ acts as 
\begin{equation}\label{S}
\mathbf{S}[f]_n=\sqrt{n}f_{n-1}+\sqrt{n+1}f_{n+1},\quad n\geq 0,
\end{equation}
with $f_{-1}=0$ and the operator rank-one operator $\mathbf{R}_1$ acts as
\begin{equation}
\mathbf{R}_1[f]_n=f_0\delta_{0,n}, \quad n\geq 0
\end{equation}
with Dirac's delta \eqref{delta}.\\
The summand $-\frac{1}{\tau}\mathbf{I}$ in \eqref{Tsplit} does not change the eigenstructure of the operator $-\ri k \mathbf{S}(k)+\frac{1}{\tau}\mathbf{R}_1$, preserving the essenial spectrum as well as the eigenfucntions, and merely induces a shift of the spectrum by $\frac{1}{\tau}$. So, to understand the spectral properties of the sum \eqref{Tsplit}, let us take a closer look at the operator \eqref{S} first.\\
First, we recall some general properties of real Jacobi operators of the form
\begin{equation}\label{J}
J[f]_n=a_nf_{n+1}+b_n+a_{n-1}f_{n-1}, \quad n\geq 1,
\end{equation}
such that $a_n>0$ and $b_n\in\mathbb{R}$, together with the initial condition 
\begin{equation}
J[f]_0=a_0f_1+b_0f_0.
\end{equation}
For a general treatment of Jacobi operators, we refer to \cite{teschl2000jacobi}. As the operator $\mathbf{S}$ is a symmetric (real) Jacobi operator, let us recall some basic properties of this class of discrete operators. which satisfies Carleman's criterion \cite[p. 49]{teschl2000jacobi}, since
\begin{equation}
\sum_{n=0}^\infty \frac{1}{\sqrt{n+1}}=\infty.
\end{equation}
Hence, it can be extended to a self-adjoint operator on $l^2(\mathbb{N})$. In particular, we have that $\sigma(\mathbf{S})\subseteq\mathbb{R}$. The spectral theory of self-adjoint Jacobi operators is closely related to the theory of orthogonal polynomials and the moment problem, cf. \cite{teschl2000jacobi}. Indeed, there exists a sequence of polynomials $\{p_n(v)\}_{n\in\mathbb{N}}$, satisfying the recurrence relation
\begin{equation}\label{Jpoly}
J[p]_n(v)=vp_n(v),\quad n\geq 0,
\end{equation}
with initial condition $p_0(v)\equiv 1$ and $p_{-1}(v)\equiv 0$, which are orthonormal with respect to the spectral measure induced by the operator $J$, cf. \cite{teschl2000jacobi}. For the operator \eqref{S}, the orthogonal polynomials produced in that fashion are just multiples of the normalized Hermite polynomials \eqref{normHermite}, namely $(2\pi)^{\frac{1}{4}}\hat{H}_n(v)$, $n\geq 0$, and the corresponding spectral measure with respect to the first basis vector is given by $e^{-\frac{v^2}{2}}\frac{dv}{\sqrt{2\pi}}$, which implies that $\sigma(\mathbf{S})=\mathbb{R}$. Let
\begin{equation}\label{GreensFunction}
G(z,n,m)=\int_{\mathbb{R}}\frac{\hat{H}_n(s)\hat{H}_m(s)}{s-z} e^{-\frac{s^2}{2}}\, ds, \quad z\in \rho(J),
\end{equation}
for $n,m\in\mathbb{N}$, denote the Green's function associated to the Jacobi operator \eqref{Jpoly} (see \cite[p.44]{teschl2000jacobi}) and let
\begin{equation}\label{diagonal}
g(z,n)=G(z,n,n), \quad z\in\rho(J),
\end{equation}
for $n\in\mathbb{N}$, be the corresponding diagonal Green's function.\\
Now, let us add the rank-one perturbation $\mathbf{R}_1$ to $\ri k\mathbf{S}$ and denote 
\begin{equation}\label{T1}
\mathbf{T}_1(k)=\mathbf{S}-\frac{1}{\ri k\tau}\mathbf{R}_1.
\end{equation}
The spectra of $\mathbf{T}(k)$ and $\mathbf{T}_1(k)$ are then related by the formula
\begin{equation}\label{spec_T}
\sigma\Big(\mathbf{T}(k)\Big)=-\ri k\sigma\Big(\mathbf{T}_1(k)\Big)-\frac{1}{\tau}.
\end{equation}
We note at this pint that the spectrum of \eqref{T1} only depends on $k$ through $k\tau$ - analogous to the the spectra analyzed in \cite{Kogelbauer2019}. By the second resolvent identity, we can write, for any $z\in\mathbb{C}\setminus \sigma(\mathbf{T}_1)$,
\begin{equation}\label{resolvent1}
(\mathbf{T}_1(k)-z)^{-1}=(\mathbf{S}-z)^{-1}+\frac{1}{\ri k\tau}\langle(\mathbf{T}_1(k)-\overline{z})^{-1}\delta_0,.\rangle(\mathbf{S}-z)^{-1}\delta_0.
\end{equation}
Writing
\begin{equation}
g_k(z)=\langle\delta_0,(\mathbf{T}_1(k)-z)^{-1}\delta_0\rangle,
\end{equation}
it follows from \eqref{resolvent1} that
\begin{equation}
g_k(z)=\frac{g(z,0)}{1-\frac{1}{\ri k\tau}g(z,0)},
\end{equation}
for the diagonal function \eqref{diagonal}. Since
\begin{equation}
(\mathbf{T}_1(k)-z)^{-1}\delta_0=\left(1-\frac{1}{\ri k\tau}g_k(z)\right)(\mathbf{S}-z)^{-1}\delta_0,
\end{equation}
it follows that
\begin{equation}\label{resolvent2}
(\mathbf{T}_1(k)-z)^{-1}=(\mathbf{S}-z)^{-1} - \frac{g(z,0)}{1-\frac{1}{\ri k\tau}g(z,0)}\langle (\mathbf{S}-\overline{z})^{-1}\delta_0,.\rangle(\mathbf{S}-z)^{-1}\delta_0.
\end{equation}
Equation \eqref{resolvent2}, which is a special case of Krein's resolvent formula \cite{kurasov2004krein}, shows that the essential spectra of $\mathbf{T}_1(k)$ and $\mathbf{S}$ agree for $k\neq 0$. Furthermore, since the spectrum of $\mathbf{S}$ is real, we see from the discrete poles of \eqref{resolvent2} that 
\begin{equation}\label{point_spec}
\sigma_{discrete}\Big(\mathbf{T}_1(k)\Big)=\left\{z\in\mathbb{C}\setminus\sigma_{ess}(\mathbf{S}): g(z,0)=\ri k \tau \right\}.
\end{equation}
For later calculations to evaluate the implicit relation in \eqref{point_spec}, we define
\begin{equation}
\erf(z)=\frac{2}{\sqrt{\pi}}\int_0^z e^{-s^2}\, ds,
\end{equation}
for $z\in\mathbb{C}$.
We calculate for $z\in\mathbb{C}\setminus\mathbb{R}$:
\begin{equation}\label{g0first}
\begin{split}
g(z,0)&=\int_{\mathbb{R}}\frac{\hat{H}^2_0(v)}{s-z}e^{-\frac{s^2}{2}}\, ds\\
&=\frac{1}{\sqrt{2\pi}}\int_{\mathbb{R}}\frac{1}{s-z}e^{-\frac{s^2}{2}}\, ds\\
\end{split}
\end{equation}
Note the different normalization of $p_0$ in \eqref{GreensFunction} and $\hat{H}_0$.
To evaluate the integral in \eqref{g0first}, we rely on the identities in \cite[p.297]{abramowitz1948handbook}. Let
\begin{equation}
w(z)=e^{-z^2}(1-\erf(-\ri z)), \quad z\in\mathbb{C},
\end{equation}
which satisfies the functional identity
\begin{equation}\label{wident}
w(-z)=2e^{-z^2}-w(z),\quad z\in\mathbb{C}.
\end{equation}
We then have that
\begin{equation}
w(z)=\frac{\ri}{\pi}\int_{\mathbb{R}}\frac{e^{-s^2}}{z-s}\, ds,\quad \Im{z}>0,
\end{equation}
and, by relation \eqref{wident}, we have for $\Im{z}<0$:
\begin{equation}
\begin{split}
\frac{\ri}{\pi}\int_{\mathbb{R}}\frac{e^{-s^2}}{z-s}\, ds&=-\frac{\ri}{\pi}\int_{\mathbb{R}}\frac{e^{-s^2}}{(-z)-s}\, ds\\
&=-w(-z)\\
&=e^{-z^2}[-1-\erf(-\ri z)].
\end{split}
\end{equation}
Consequently, we obtain
\begin{equation}
\begin{split}
\int_{\mathbb{R}}\frac{1}{s-z}e^{-\frac{s^2}{2}}\, ds&=\int_{\mathbb{R}}\frac{e^{-s^2}}{s-\frac{z}{\sqrt{2}}}\, ds\\
&=\ri\pi\frac{\ri}{\pi}\int_{\mathbb{R}}\frac{e^{-s^2}}{\frac{z}{\sqrt{2}}-s}\, ds\\
&=\begin{cases}
\ri\pi e^{-\frac{z^2}{2}}\left[1-\erf\left(\frac{-\ri z}{\sqrt{2}}\right)\right],&\quad\text{ if } \Im{z}>0,\\
\ri \pi e^{-\frac{z^2}{2}}\left[-1-\erf\left(\frac{-\ri z}{\sqrt{2}}\right)\right],&\quad\text{ if } \Im{z}<0,
\end{cases}
\end{split}
\end{equation}
where in the first step, we have re-scaled $s\mapsto \sqrt{2}s$ in the integral.
Written more compactly, we arrive at
\begin{equation}\label{g0}
g(z,0)=\frac{\ri\pi}{\sqrt{2\pi}} e^{-\frac{z^2}{2}}\left[\sign(\Im{z})-\erf\left(\frac{-\ri z}{\sqrt{2}}\right)\right].
\end{equation}
The function \eqref{g0} maps imaginary numbers to imaginary numbers. In particular, for
\begin{equation}
z=\sqrt{2}\ri x, 
\end{equation}
equation \eqref{point_spec} is equivalent to the real equation
\begin{equation}\label{point_eq}
\sqrt{\frac{\pi}{2}} e^{x^2}\left(\sign (x)-\erf(x)\right)=k\tau.
\end{equation}
The left-hand side of equation \eqref{point_eq} up to a factor is depicted in Figure \ref{Green}.

\begin{figure}
	\includegraphics[scale=0.4]{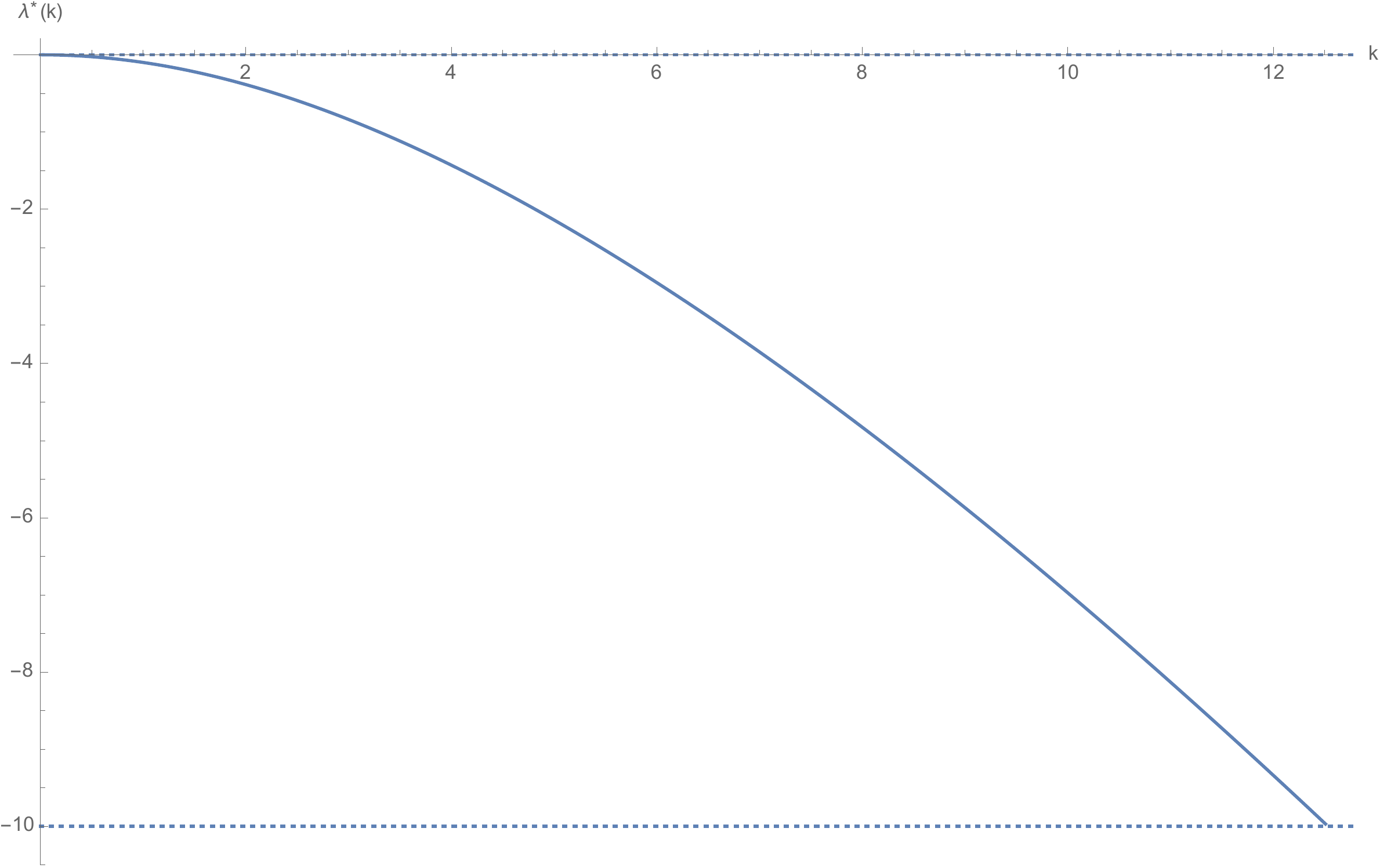}\caption{The eigenvalue $\lambda^*$ in dependence of the wave number $k$ for $\tau=0.1$. The critical wave number is at $k_{crit}\approx 12.5331$.}
	\label{lambda*}
\end{figure}

\begin{figure}
	\includegraphics[scale=0.7]{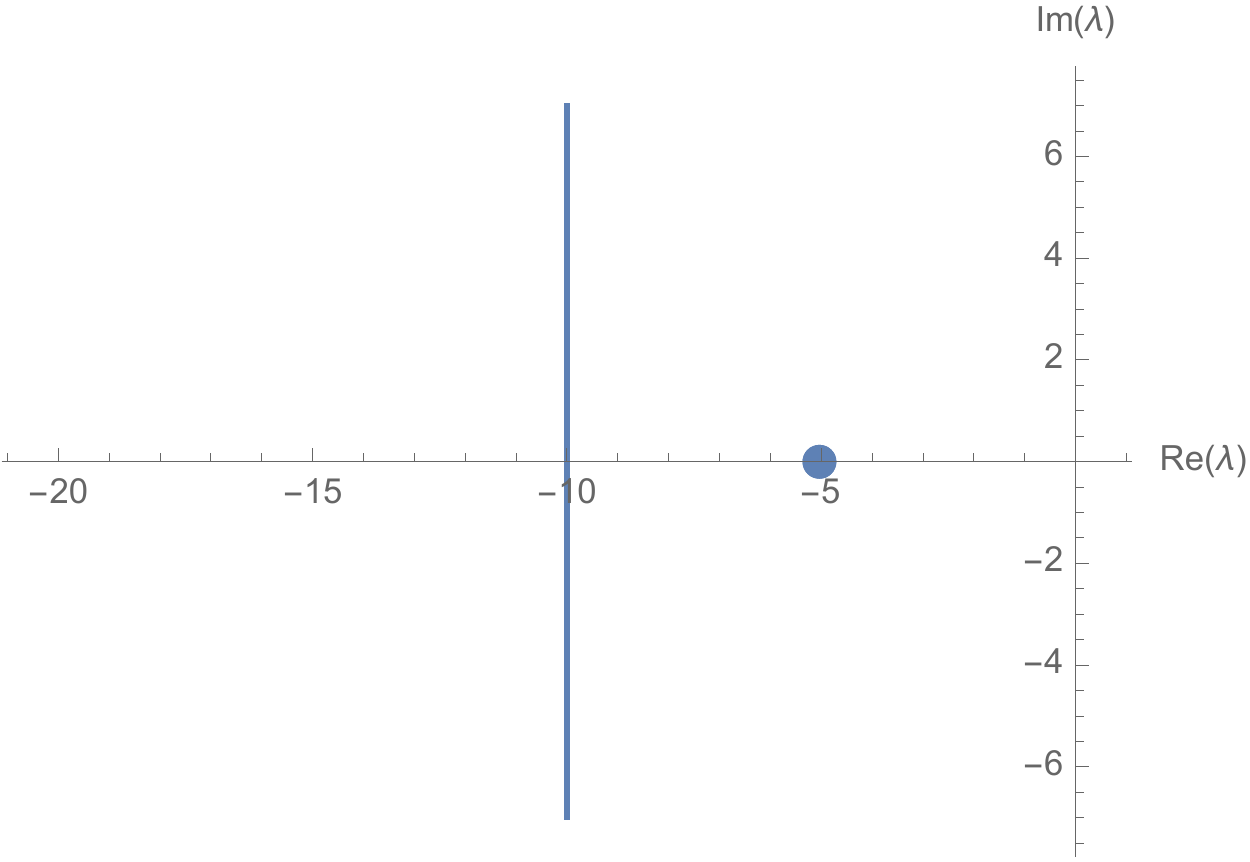}\caption{Typical spectrum of the operator $\mathbf{T}(k)$ for $0<|k|<k_{crit}$.}
	\label{spectrum_T}
\end{figure}

 Since the diagonal Green's function $\tilde{g}(x):=-\ri g(\ri x,0)$ is bounded, we have that the discrete spectrum of $\mathbf{T}$ is non-empty only for a certain range $[-k_{crit},k_{crit}]$ of wave numbers, excluding $k=0$. From \eqref{point_eq}, it follows that the critical wave number is given by 
 \begin{equation}
k_{crit}=\sqrt{\frac{\pi}{2}}\frac{1}{\tau}.
 \end{equation} 
 Let $x^{*}(k\tau)$ denote the unique solution to equation \eqref{point_eq} for $0<|k|<k_{crit}$ and let $\lambda^*(k,\tau)$ denote the unique eigenvalue of $\mathbf{T}(k)$ for $0<|k|<k_{crit}$. From \eqref{spec_T}, it follows that 
 \begin{equation}\label{deflambda*}
 \lambda^*(k,\tau)=-\frac{1}{\tau}-\ri\sqrt{2} k\ri x^*(k\tau)=-\frac{1}{\tau}+\sqrt{2}k x^*(k\tau). 
 \end{equation}
Figure \ref{lambda*} shows the eigenvalue $\lambda^*(k,\tau)$ for $0<|k|<k_{crit}$ and $\tau=0.1$. The typical spectrum of $\mathbf{T}(k)$ for $0<|k|<k_{crit}$ is depicted in Figure \ref{spectrum_T}.

To complete the spectral analysis of the operator $\mathbf{T}(k)$, let us describe the eigenvector associated to $\lambda^*(k,\tau)$ in more detail. The eigenvector associated to the eigenvalue $\lambda^*(k,\tau)$ is the unique solution to system \eqref{resolvent} for $\lambda=\lambda^*(k,\tau)$ and $\eta_n=0,\quad n\geq0$. By setting $\hat{f}_n=\sqrt{n!}g_n$ and division by $-\ri k\sqrt{n!}$, system \eqref{resolvent} becomes
\begin{equation}\label{resolventg}
\begin{split}
g_{n-1}+\frac{1}{\ri k}\left(\frac{1}{\tau}+\lambda^*(k,\tau)\right)g_n+ (n+1)g_{n+1}&=0,\quad n\geq 1,\\
g_1+\frac{\lambda^*(k,\tau)}{\ri k} g_0&=0.
\end{split}
\end{equation}
To simplify notation, we set
\begin{equation}
\mu(k):=\frac{1}{\ri k}\left(\frac{1}{\tau}+\lambda^*(k)\right).
\end{equation}
In order to solve system \eqref{resolventg}, define the generating function of the sequence $(g_0,g_1,g_2,...)$ as
\begin{equation}
\Gamma(z)=\sum_{n=0}^\infty g_n z^n.
\end{equation}
Since 
\begin{equation}
z\Gamma(z)=\sum_{n=0}^{\infty} g_nz^{n+1}=\sum_{n=0}^{\infty}g_{n-1}z^n,
\end{equation}
where we have set $g_{-1}=0$, and since
\begin{equation}
\Gamma'(z)=\sum_{n=0}^{\infty}ng_n z^{n-1}=\sum_{n=0}^{\infty}(n+1)g_{n+1} z^{n},
\end{equation}
the equation
\begin{equation}\label{eqGamma}
z\Gamma(z)+\mu(k)\Gamma(z)+\Gamma'(z)=\left(\mu(k)-\frac{\lambda^*(k)}{\ri k}\right)g_0,
\end{equation}
is equivalent to system \eqref{resolventg}. Note that equation \eqref{eqGamma} also reproduces the zeroth-order equation in system \eqref{resolventg}. Setting 
\begin{equation}
\sigma(k):=\left(\mu(k)-\frac{\lambda^*(k)}{\ri k}\right),
\end{equation}
the solution to equation \eqref{eqGamma} is given by
\begin{equation}
\Gamma(z)=e^{-\frac{1}{2}z(2\mu(k)+z)}\left(c_0+\sqrt{2}\sigma(k)g_0e^{-\frac{1}{2}\mu^2(k)}\erf\left(\frac{z+\mu(k)}{\sqrt{2}}\right)\right),
\end{equation}
for some constant $c_0\in\mathbb{C}$. To match the condition $\Gamma(0)=g_0$, we have to set
\begin{equation}
c_0=\left(1-\sqrt{2}\sigma(k)e^{-\frac{1}{2}\mu^2(k)}\erf\left(\frac{\mu(k)}{\sqrt{2}}\right)\right)g_0.
\end{equation}
This is consistent with the fact that an eigenvector is unique up to re-scaling. Defining Dawson's function
\begin{equation}
D_{+}(x)=2^{-x^2}\int_0^{x}e^{y^2}\, dy,
\end{equation}
and remembering that $g_0=\hat{f}_0$, we can write the solution $\Gamma$ more compactly as
\begin{equation}
\Gamma(z)=e^{-\frac{1}{2}z(2\mu(k)+z)}\left(1-\sqrt{2}\sigma(k)D_{+}\left(\frac{\mu(k)}{\sqrt{2}}\right)\right)+\sqrt{2}\sigma(k)D_{+}\left(\frac{\mu(k)+z}{\sqrt{2}}\right)\hat{f}_0.
\end{equation}
The eigenvector associated to the eigenvalue $\lambda^*(k)$ is then given as
\begin{small}
\begin{equation}\label{f*}
\hat{f}^*_n(k)=\sqrt{n!}\left.\frac{d^n}{dz^n}\right|_{z=0}\left[ e^{-\frac{1}{2}z(2\mu(k)+z)}\left(1-\sqrt{2}\sigma(k)D_{+}\left(\frac{\mu(k)}{\sqrt{2}}\right)\right)+\sqrt{2}\sigma(k)D_{+}\left(\frac{\mu(k)+z}{\sqrt{2}}\right) \right]\hat{f}_0(k).
\end{equation}
\end{small}
Note that we can choose a different value of $\hat{f}_0$ for each wave number $k$.\\

\section{Hydrodynamic Closure, Asymptotic Expansion of $\lambda^*(k)$ and Comparison to CE expansion}

In this section, we interpret the hydrodynamic closure obtained in the previous section physically and compare its asymptotic expansion (for small relaxation time $\tau$) to the Chapman--Enskog expansion.

\subsection{Hydrodynamic Closure}
In the previous section, we have found a value $\lambda^*(k,\tau)$ as defined in \eqref{lambda*} and a sequence of moments $\mathbf{F}^*(k)=(\hat{f}_0(k),\hat{f}_1^*(k),\hat{f}_2^*(k),...)$, for some $k\mapsto\hat{f}_0(k)\in\mathbb{C}$, as defined in \eqref{f*} such that
\begin{equation}
\mathbf{T}(k)\mathbf{F}^*(k)=\lambda^*(k,\tau)\mathbf{F}^*(k).
\end{equation}
This eigenvalue constitutes a slow mode of the overall dissipative system, i.e., the spectrum has negative real part and the remaining part of the spectrum (excluding the eigenvalue) has a real part strictly less than $\Re\lambda^*(k,\tau)$. This holds as long as $0<|k|<k_{crit}$. Therefore, the overall dynamics generated by the operator $\mathbf{T}(k)$ approach the projected dynamics on the eigenvector \eqref{f*} exponentially fast. For $k=0$, the eigenvector $(1,0,0,...)$ actually defines a center-manifold on which the dynamics are trivial (base state).\\
If we now assume that $\mathbf{F}(t)=\mathbf{F}^{*}(k,t)=\mathbf{F}^{*}_{coef}(k)\hat{f}_0(k,t)$, we find that the operator equation \eqref{eqF} reduces to
\begin{equation}\label{closedk}
\frac{\partial}{\partial t}\hat{f}_0(k,t)=\lambda^*(k,\tau)\hat{f}_0(k,t).
\end{equation}
Taking an inverse Fourier transform of the hydrodynamic closure in frequency space \eqref{closedk}, we obtain
\begin{equation}\label{closedx}
\frac{\partial}{\partial t}f_0(x,t)=\frac{1}{2\pi}\int_{\mathbb{R}}\lambda^*(k,\tau)\hat{f}_0(k,t) e^{\ri k x}\, dk.
\end{equation}
Formula \eqref{closedx} defines a non-local hydrodynamic closure of system \eqref{maineq} independent on the size of the relaxation time $\tau$.

\subsection{Asymptotic Expansion of $\lambda^*(k)$ and Comparison to CE expansion}
To find the asymptotic behavior of $\lambda^*(\tau,k)$ for small values of $\tau$, let us define
\begin{equation}
\phi(y)=e^{x^2}\left(1-\erf(x)\right)|_{x=\frac{1}{y}}.
\end{equation}
Since $\tilde{g}(x)$ is an odd function, it suffices to analyze the properties of the positive branch. Due to the decay properties of $\tilde{g}(x)$ for $x\to\pm\infty$, the function $\phi(y)$ is analytic around $y=0$ (here, we neglect the jump discontinuity in the definition of $\tilde{g}$). Also, we find that
\begin{equation}
\begin{split}
\phi(0)&=0,\\
\phi'(0)&=\frac{1}{\sqrt{\pi}}.
\end{split}
\end{equation}
Let
\begin{equation}
\varepsilon=\sqrt{\frac{2}{\pi}}k\tau.
\end{equation}
From the Lagrange--B\"urmann inversion formula \cite{abramowitz1948handbook}, we can write the solution to the equation $\phi(y^*(\varepsilon))=\varepsilon$ as
\begin{equation}
y^*(\varepsilon)=\sum_{n=1}^{\infty}\left(\left.\frac{d^{n-1}}{dy^{n-1}}\right|_{y=0}\left(\frac{y}{\phi(y)}\right)^{n}\right)\frac{\varepsilon^n}{n!}.
\end{equation}
The first few terms in the expansion are given as
\begin{equation}
y^*(\varepsilon)=\sqrt{\pi}\varepsilon+\frac{\pi^{\frac{3}{2}}}{2}\varepsilon^3+\frac{3\pi^{\frac{7}{2}}}{8}\varepsilon^7+\mathcal{O}(\varepsilon^9).
\end{equation}
From equation \eqref{deflambda*} and the definition of $\varepsilon$, we have that
\begin{equation}\label{asymptotic}
\begin{split}
\lambda^*(k,\tau)&=-\frac{1}{\tau}+\sqrt{2}kx^*(\varepsilon)\\
&=k\left(-\frac{1}{\sqrt{\frac{\pi}{2}}\varepsilon}+\frac{\sqrt{2}}{y^*(\varepsilon)}\right)\\
&=k\left(\frac{-\left(\sqrt{\pi}\varepsilon+\frac{\pi^{\frac{3}{2}}}{2}\varepsilon^3+\frac{3\pi^{\frac{7}{2}}}{8}\varepsilon^7+\mathcal{O}(\varepsilon^9)\right)+\sqrt{\pi}\varepsilon}{\sqrt{\frac{\pi}{2}}\varepsilon(\sqrt{\pi}\varepsilon+\frac{\pi^{\frac{3}{2}}}{2}\varepsilon^3+\frac{3\pi^{\frac{7}{2}}}{8}\varepsilon^7+\mathcal{O}(\varepsilon^9)) }\right)\\
&=k\left(\frac{-\frac{\pi^{\frac{3}{2}}}{2}\varepsilon-\frac{3\pi^{\frac{7}{2}}}{8}\varepsilon^5+\mathcal{O}(\varepsilon^7)}{\sqrt{\frac{\pi}{2}}(\sqrt{\pi}+\frac{\pi^{\frac{3}{2}}}{2}\varepsilon^2+\frac{3\pi^{\frac{7}{2}}}{8}\varepsilon^6+\mathcal{O}(\varepsilon^8))}\right)\\
&=-\varepsilon k\sqrt{2\pi}\left(\frac{4+3\pi^2\varepsilon^4+\mathcal{O}(\varepsilon^6)}{8+4\pi\varepsilon^2+3\pi^3\varepsilon^6+\mathcal{O}(\varepsilon^8)}\right)\\
&=\frac{-k\varepsilon}{2}\sqrt{\frac{\pi}{2}}[2-\pi\varepsilon^2+2\pi^2\varepsilon^4+\mathcal{O}(\varepsilon^6)]\\
&=-k^2\tau+k^4\tau^3-4k^6\tau^5+\mathcal{O}(\tau^7,k^8).
\end{split}
\end{equation}
At leading order, we find that the non-local hydrodynamic closure \eqref{closedx} becomes the heat equation
\begin{equation}\label{order1}
\frac{\partial f}{\partial t}=\tau\frac{\partial^2f}{\partial x^2},
\end{equation}
while at order $\tau^3$, equation \eqref{closedx} becomes
\begin{equation}\label{order3}
\frac{\partial f}{\partial t}=\tau\frac{\partial^2f}{\partial x^2}+\tau^3\frac{\partial^4f}{\partial x^4}.
\end{equation}
\begin{remark}
	The instability in the dispersion relation (wave number) is due to the coupling of the wave number $k$ to the relaxation time $\tau$. A (formal) asymptotic expansion in $\tau$ necessarily leads to a polynomial - hence unbounded - approximation of the closure relation, while the multiplier in the full (non-local) hydrodynamic closure \eqref{closedk} remains bounded in wave space for all $k\in\mathbb{R}$ with $0<|k|<k_{crit}$.
\end{remark}
In order to compare equations \eqref{order1} and \eqref{order3} with the CE series, we define the momenta
\begin{equation}
M_l(x,t):=\int_{\mathbb{R}}v^lf(v,x,t)\, dv, \quad l\geq 0.
\end{equation}
Equation \eqref{maineq} can be rewritten as an infinite-dimensional moment system
\begin{equation}\label{moment}
\frac{\partial M_{l}}{\partial t}=-\frac{\partial M_{l+1}}{\partial x} -\frac{1}{\tau}M_l+\frac{1}{
\tau}M_l^{eq}, \quad l\geq 0,
\end{equation}
where, for $l \geq 0$, 
\begin{equation}
\begin{split}M_l^{eq}(x,t)&=\frac{1}{\sqrt{2\pi}}\int_{\mathbb{R}}\rho(x,t)e^{-\frac{v^2}{2}}v^l\, dv\\
&=\frac{\rho}{\sqrt{2\pi}}\left(\left[-e^{-\frac{v^2}{2}}v^{l-1}\right]_{-\infty}^\infty+(l-1)\int_{\mathbb{R}}e^{-\frac{v^2}{2}}v^{l-2}\, dv\right)\\
&=\begin{cases}(l-1)!!\rho(x,t) &\text{ for } l \text{ even}\\0 &\text{ for } l \text{ odd },
\end{cases}
\end{split}
\end{equation}
since $\int_{\mathbb{R}}e^{-\frac{v^2}{2}}\, dv=\sqrt{2\pi}$ and with the convention that $(-1)!!=1$.
Applying the Fourier transform to equation \eqref{moment}, we obtain
\begin{equation}
\frac{\partial \hat{M}_{l}}{\partial t}=-\ri k \hat{M}_{l+1}-\frac{1}{\tau}\hat{M}_l+\frac{1}{\tau}\frac{1+(-1)^l}{2}(l-1)!!\hat{M}_0,
\end{equation}
which can be written as
\begin{equation}\label{eqM}
\hat{\mathbf{M}}_t=\mathbf{A}(k,\tau)\hat{\mathbf{M}},
\end{equation}
for the sequence $\hat{\mathbf{M}}=\{\hat{M}_0,\hat{M}_1,\hat{M}_2,...\}$ and the infinite matrix
\begin{equation}\label{A}
\mathbf{A}(k,\tau)=\left(\begin{matrix}0  & -\ri k & 0 & 0 & 0 & 0 &\ldots \\0 & -\frac{1}{\tau} & -\ri k & 0 & 0 & 0 &\dots \\1 & 0 & -\frac{1}{\tau} & -\ri k  & 0 & 0 & \dots\\0 & 0 & 0 & -\frac{1}{\tau} & -\ri k & 0 & \ldots\\3 & 0 & 0 & 0 & -\frac{1}{\tau} & -\ri k & \dots\\\vdots & \vdots & \vdots & \vdots & \vdots & \vdots & \ddots\end{matrix}\right).
\end{equation}
We can write system \eqref{moment} equivalently as
\begin{equation}\label{moment2}
\begin{split}
\frac{\partial \hat{\rho}}{\partial t}&= -\ri k \hat{M}_1,\\
\frac{\partial\tilde{\mathbf{M}}}{\partial t}&=-\ri k \mathbf{\Sigma}\tilde{\mathbf{M}}-\frac{1}{\tau}(\tilde{\mathbf{M}}-\mathbf{M}^{eq}),
\end{split}
\end{equation}
with $\tilde{\mathbf{M}}=(\hat{M}_1,\hat{M}_2,...)$, $\hat{\mathbf{M}}^{eq}=(\hat{M}_1^{eq},\hat{M}_2^{eq},...)$ and with the shift operator
\begin{equation}
\mathbf{\Sigma}=\left(\begin{matrix}0  & 1 & 0 & 0 & 0 & 0 &\ldots \\0 & 0 & 1 & 0 & 0 & 0 &\dots \\0 & 0 & 0  & 1 & 0 & 0 & \dots\\0 & 0 & 0 & 0 & 1 & 0 & \ldots\\0 & 0 & 0 & 0 & 0& 1& \dots\\\vdots & \vdots & \vdots & \vdots & \vdots & \vdots & \ddots\end{matrix}\right).
\end{equation}
Assuming that the relaxation time $\tau$ is small, the Chapman--Enskog series for system \eqref{maineq} is a (formal) Taylor expansion of the Fourier transform of the higher moments,
\begin{equation}\label{CEexpansion}
\hat{M}_n(k,t;\tau)=\sum_{j=0}^\infty m_{n,j}(k,t)\tau^j,\quad n\geq 1.
\end{equation}
Now, we can plug expansion \eqref{CEexpansion} into equation \eqref{moment2} to obtain that
\begin{equation}\label{m}
\frac{\partial }{\partial t}m_{n,j}=-\ri k m_{n+1,j}-m_{n,j+1},\quad n\geq 1, \quad j\geq 0,
\end{equation}
provided we set
\begin{equation}
m_{n,0}=\frac{1+(-1)^n}{2}(l-1)!!\rho, \quad n\geq 1.
\end{equation}
We can solve equation \eqref{m} recursively to obtain
\begin{equation}
m_{1,1}=-\ri k \rho, \quad m_{1,2}=0, \quad m_{1,3}=k^4,
\end{equation}
implying that
\begin{equation}
\frac{\partial \hat{\rho}}{\partial t}=(-k^2\tau+k^4 \tau^3)\hat{\rho} +\mathcal{O}(\tau^4),
\end{equation}
which is consistent with the asymptotic expansion \eqref{asymptotic}.

\section{Conclusion and Further Perspectives}
We have provided an exact non-local hydrodynamic closure for the one-dimensional kinetic equation independent of the size of the relaxation time. The obtained evolution equation is consistent with the CE series approximation and implies an explicit formula for the critical wave number.\\
The current analysis shows that a dynamical systems point of view might prove useful in the further investigation of the hydrodynamic closure. In particular, the existence and qualitative behavior of hydrodynamic manifolds for nonlinear model equations, based on invariant manifold techniques, is  a promising way to make further progress in this field. 

\section{Acknowledgments}
The author would like to thank Gerald Teschl for very useful discussions on Jacobi operators and the spectral theory of rank-one perturbations. The author would like to thank Ilya Karlin for pointing out this direction of research and several useful discussions and comments in the context of statistical physics and the CE series.

\bibliographystyle{abbrv}
\bibliography{/Users/floriankogelbauer/Dropbox/Bibs/DynamicalSystems}
 \end{document}